# Dynamics of popstar record sales on phonographic market - stochastic model

Andrzej Jarynowski

Smoluchowski Institute, Jagiellonian University, Cracow, Poland

Department of Sociology, Stockholm University, Sweden

CIOP, National Research Institute, Warsaw, Poland

Andrzej Buda

Institute of Nuclear Physics, Cracow, Poland

Wydawnictwo Niezalezne, Wroclaw, Poland

**Abstract**

We investigate weekly record sales of the world's most popular 30 artists (2003-2013). Time series of sales have non-trivial kind of memory (anticorrelations, strong seasonality and constant autocorrelation decay within 120 weeks). Amount of artists record sales are usually the highest in the first week after premiere of their brand new records and then decrease to fluctuate around zero till next album release. We model such a behavior by discrete mean-reverting geometric jump diffusion (MRGJD) and Markov regime switching mechanism (MRS) between the base and the promotion regimes. We can built up the evidence through such a toy model that quantifies linear and nonlinear dynamical components (with stationary and nonstationary parameters set), and measure local divergence of the system with collective behavior phenomena. We find special kind of disagreement between model and data for Christmas time due to unusual shopping behavior. Analogies to earthquakes, product life-cycles, and energy markets will also be discussed.

89.65.Gh, 89.65.-s, 89.75.-k, 05.10.Gg, 05.40.-a, 05.40.Fb, 89.65.Gh, 02.50.Ga, 47.27.tb

## 1  Phonographic Market of best selling artists

The phonographic market already known as an example of complex system [1, 2], is much more predictable than finance market [3, 4]. 80% of the market is dominated by four companies (Universal, EMI, Sony BMG and Warner Bros.). All the best-selling artists belong to these major labels. There are many evidences (like changing premiere date, decisions of continuing or stopping promotion, etc) that market is controlled by them. Only sudden deaths could kick off the whole system (as it has been showed in our previous article [1, 3]). That is why we decided to model the dynamics of record sales because phonographic market is a commodity market where artists are products and music fans are customers.

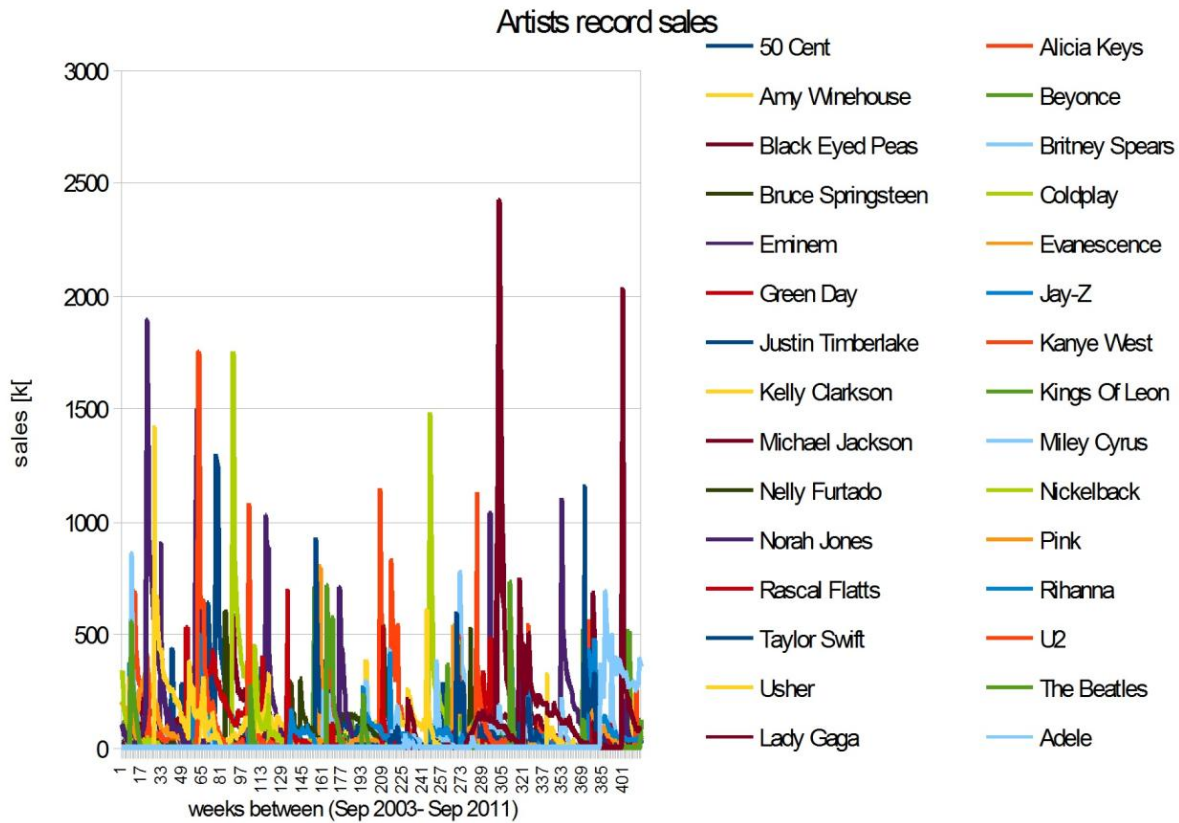

Figure 1: Chosen artists cohort and their weekly record sales

In our analysis we have chosen 30 best selling artists due to homogeneity in market strategies and data availability. Only a few researches were provided for understanding successes or failures of an artist, mainly in economic frameworks [8, 9] of product life-cycle or innovation spread. There were also culture studies focused on ethnology and customers decisions [10, 11]. Unfortunately, quantitative methods are rarely used in these analyzes. If it is applied, linear regression dominates [12]. A slightly less well known but related superstar phenomenon [5] is even more interesting from complex system perspective due to herd effect and criticality of customers decision. In this work, we model the challenges for artists and record labels staying behind their market decisions in ultimate sales game. We have focused on artists popularity that is measured by record sales (see Fig. 1). Unfortunately, exact record sales (day by day, or within all geographic areas) have been officially unavailable yet. However, IFPI (International Federation of the Phonographic Industry) publishes weekly chart of the 200 world's best selling albums with precision of 100 units. These data are the source of our time series[6]. Unfortunately, only a sales value is given for albums on the list, so any sales below some threshold (which also vary in time very much) are approximated by zero. However, under the 200th place of the chart record, sales are so low. Thus, it is possible to claim that these record sales (a few thousand copies) are equal to zero because the chart-topping albums sell 500 000 - 1000 000 units per week. The aim of our article is to fill in the gap between phonographic market analyzes and complex system approach [7] by applying discrete stochastic process simulations (MRGJD+MRS) developed previously in field of earthquakes and energy markets. In the MRGJD case, we add to the geometric Vasicek (MRGD) process the jumps (described by fat-tailed distribution) with given probability. In MRS approach, we introduce a transition matrix of switching probabilities (based on an empirical observation of the system) between the base or the record promotion regime.

## 2 Some rules best selling artists act and market statistics

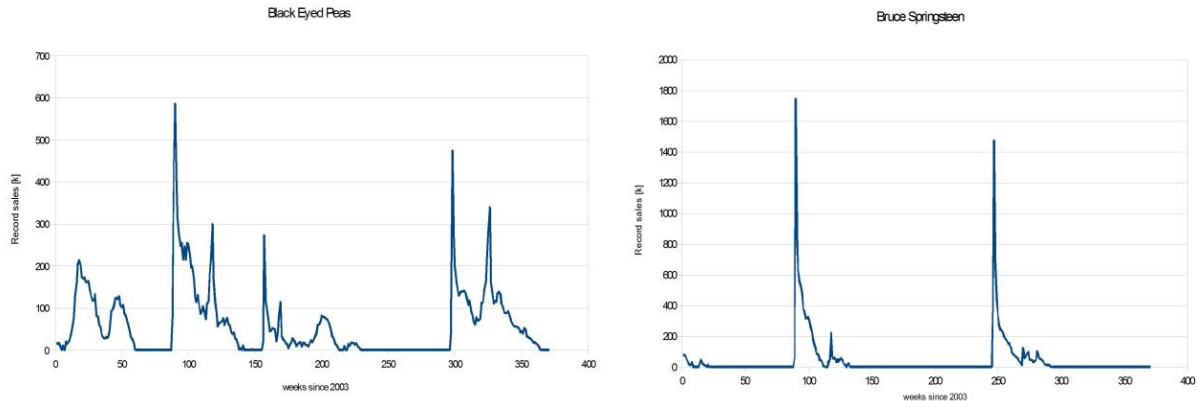

Figure 2: Examples of artists sale trajectories in time: Bruce Springsteen [left], Black Eyed Peas [right]

Phonographic market could be represented as a set of players (artist) managed by their labels, who tried to sell product to consumers. As the size of market is limited, labels with their artists are trying to establish optimal strategy. In our perspective an artist is a product. They can play with many variables like data of album release and amount of invest (in terms of insensitivity and duration) in an album promotion. Empirical observations (Fig. 2) show that album sales of an artist reaches a maximum exactly during the week of the release its new album in rapid and spiky way (such as spike are often observed in price on energy markets [13]. "Big Four" companies (EMI, Universal, Warner, Sony BMG) with a view to maximum sales will, however, set release dates plate from above, even in agreement with each other, often months in advance. A premiere of the album is often accompanied by a promotional single, which presence in the media raises the sale. Therefore, it is easier to control, since just after the release decline in sales is observed, similar to the exponential one (Figure 2). Of course, labels want to prevent it, so releasing more singles from the album stops for a while a decrease of its sales, which also can be seen (Figure 2) in the shape of the decay after premieres.

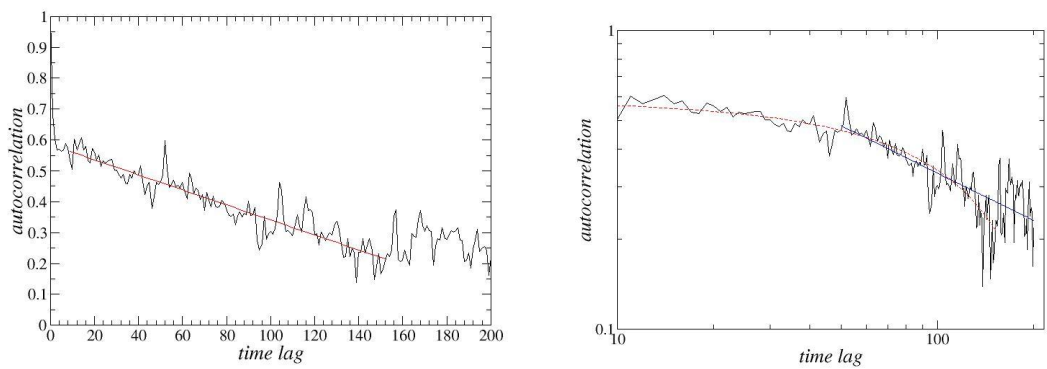

Figure 3: Autocorelation function for record sale on linear [left] and half-logarithmic [right]

A similar phenomenon: a sharp peak and spiky mean reverting process with secondary shocks (so-called aftershocks, which are analogous to the presence of the following CD singles boost sales) are also observed in earthquakes [14]. Moreover, so-called pre-shocks are detected just before the main earthquake. Those shocks of low intensity appear (or not) before the main peak. A similar phenomenon is coming also in the music market, as a result of promotion launching. Before the delivery of a new material to the market, sales of all previous titles for the artist could increase. This phenomenon occurs only in some of the artists (e.g., it does not occur with newcomers at all), as well as only few earthquakes are preceded by a pre-shock shocks.

Due to the complexity and uncertainties in the power grid, electricity prices are also caring with spikes, which may be tens or even hundreds of times higher than the normal price. Such spikes in energy markets represent demand shifts in the presence of inelastic supply or strategic withholding by suppliers. The main difference between earthquakes and record sales in opposition to spikes in energy market is a relaxation time from an exited state. As in energy market, seasonality plays a big role in record sale, because both sectors depend strongly on human behavioral cycles (Fig. 4). This energy, and phonographic markers are found to be well described by an anti-persistent (mean-reverting) walk with characteristic Hurst exponent of 0.41 [15] for energy and 0.39 [3] for phonographic market (it is not the case for earthquakes, where Hurst exponent is mainly above 0.5 [16]). For electricity spot price autocorrelation falls into the confidence interval of noise very quickly (within few hours, while hour is usually the minimal trading unit) with an additional strong 7-day (week) dependence [17]. However, autocorrelation decays much slower (up to 120 weeks) in phonographic market with small 52 weeks (year) dependence (Fig. 3). For earthquakes autocorrelation decay [18] can vary from very short (hours) to very long (weeks) decays without a special kind of seasonality.

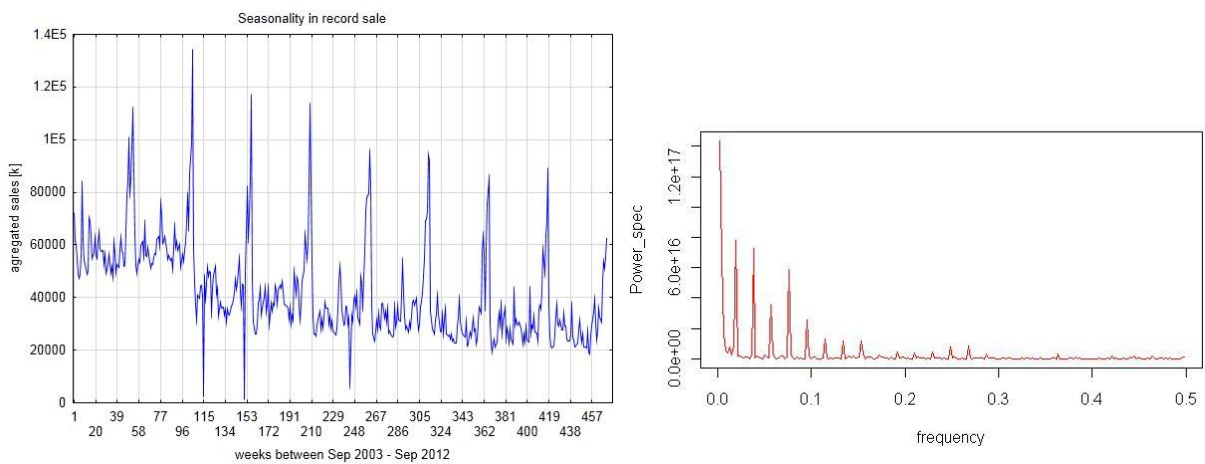

Figure 4: Seasonality. Aggregated record sale [left] and its Fourier transform [right]. Characteristic period 52 weeks (frequency 1/52) can be observed

Even a different underlying mechanism, all mentioned phenomena can be modeled statistically in a similar way [14, 13].

## 3 Empirical observation of marketing strategies

### 3.1 Album release

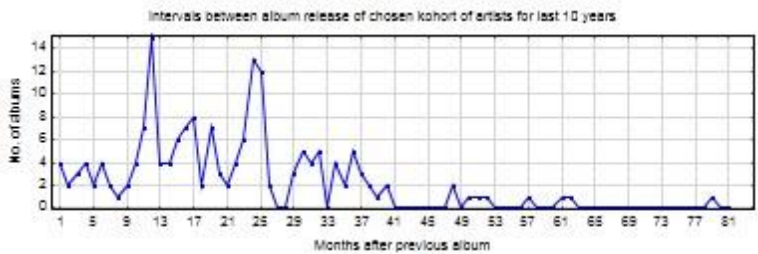

Figure 5: Intervals between albums

In opposite to earthquakes or energy price crises, release date - and therefore the peak sales or "major shock" - are set arbitrary by the record label, several weeks before the premiere. In general nowadays, the top artists release albums every 2 years (Fig. 5). This interval seems to be a natural duration of a creative process for a typical studio album length [19]. As a result, labels appear in the meantime greatest hits compilations of the artist, or his live recordings, usually at the end of the year, to shorten the waiting time for the new album. In this way, labels adapt to the seasonal demand. Every year we observe that albums released late Autumn/early Winter or on a special occasion (e.g. Christmas, various anniversaries, etc.) enjoy great popularity [11]. On the other hand, occasional albums do not have such an enormous peak on the release week and they sell much less in competition to studio albums. From intervals statistic (Fig. 5), we find the main peak between weeks 122−130 and second around 52 (exactly one year). From premier distribution (Fig. 6), it turns out that the record companies are planning to increase sales by releasing albums of celebrities usually in the months of September to November. In Autumn time, the observed intensity of the premier is 3 times higher than in the rest of the year, where their distribution is very uniform (Fig. 6). Please remark, that such an increase of album release is observed only for superstars and for all artists in data set only post Christmas/New Year Eve decrease is observed. It can be explain that young artists prefer to publish their records out of the hot period to be recognized and do not disappear in the celebrities regime. Planning date of premiere hone as to do with the fact that a few weeks before Christmas new costumers join the market. Those people buy only the 'shine' album usually as a gift, and their knowledge of artists is limited to the most popular celebrities.

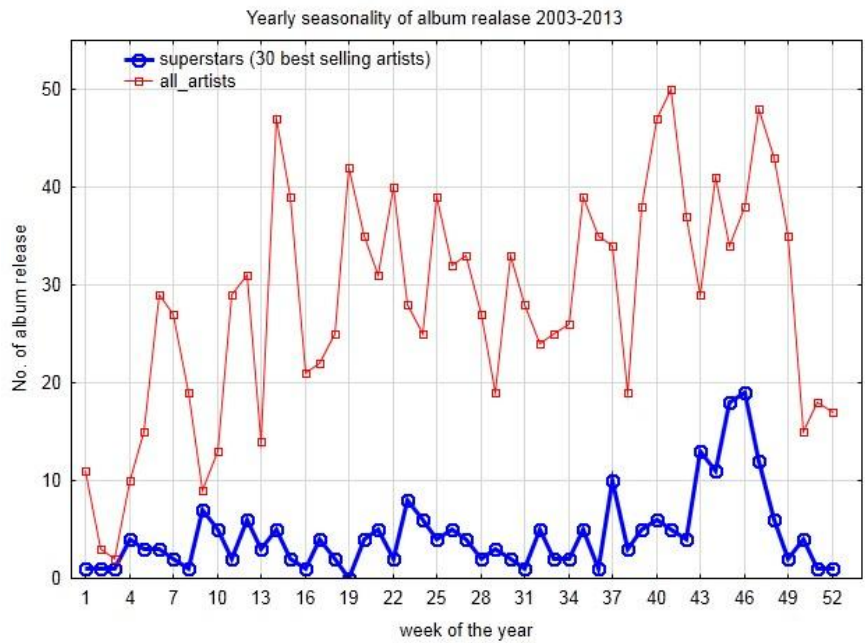

Figure 6: Album release seasonality for cohort chosen in study (30 best selling artist) and additionally all artists

## 3.2 Album promotion and singles release

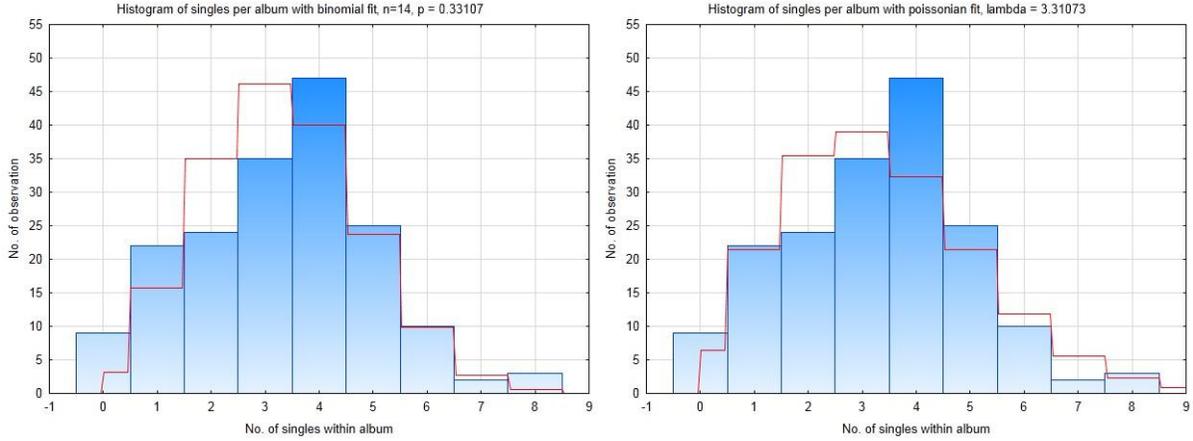

Figure 7: Histograms of number of singles in album with binomial [left] and poissonian [right] fit

Another important factor is the promotion, which goes to radio stations, music, television and the Internet. In practice, if the first single will be a hit, the label promotes album further - as long as there is a demand. When it turns into a failed single hit, given album falls from the album list and the promotion stops. Additional property is the duration of the promotion, which in practice is equal to the number of weeks the album of an artist spends on the sales charts since launch. We decided to link a duration of a promotion with a number of singles (Fig. 7), while some artists are not promoted at all, others hit more than 10 singles, resulting in sales growth of stem albums and extending their presence on the list for several years.

## 3.3 Album recognition

Of course, promotion is not only hits, but also tours and a number of other marketing activities. Especially because of the presence of the lists one can determine another parameter - the exceptional quality of the product. For well-known artists, every third of studio albums does not decay exponentially, but stepwise due to unpredictable popularity. Those albums seems to reach wider consumer range as it was initially targeted.

## 4 Model: Stationary case

We focus on Markov regime-switching (MRS) models, which seem to be a natural candidate for modeling the spiky, non-linear behavior of record's sale. Apart from a noisy base regime, an exited (promotion of record) regime was introduced. Both regimes are described by a mean-reverting geometric Random Walk (MRGRW) - type called also mean-reverting geometric Brownian Motion (MRGBM). However, for the exited (record promotion) regime(s) a number of specifications apply. The Brownian motion $X$ is responsible for relatively small (proportional to $s$ multiplied by current value of the precess ) fluctuations around the long-term mean $a/b$ (1).

$$dX_t = (a - bX_t)dt + sX_t dW_t \quad (1)$$

Discrete version $dt=1$week (Euler extension):

$$X(t+1) = X(t) + a - bX(t) + sX(t)N(0,1) \quad (2)$$

where N(0,1) is a random variable from normalized normal distribution.
Markov regime-switching (MRS) mechanism describes switching between base 1 and promotion 2 regimes. We can define transition matrix which contains probabilities of switching between regime $i$ to $j$. Regime switch can accrue every time point with resolution already mentioned above: $dt=1$week.

$$\{q_{ij}\}= \begin{bmatrix} q_{11} & q_{12} \\ q_{21} & q_{22} \end{bmatrix} = \begin{bmatrix} 1-q_{12} & q_{12} \\ 1-q_{22} & q_{22} \end{bmatrix} \quad (3)$$

We can understand for example, that time in which process stays in state 2 is random variable from exponential distribution with mean $1/q22$.
Equation for both regimes should look like:

$$dX_t = (a - bX_t)dt + sX_t dW_t + j(t)(dQ_t + dq_t) \quad (4)$$

where:

$$j(t) = \begin{cases} 1 & \text{if state } 1-> 2 \\ 0 & \text{otherwise} \end{cases}$$

Such an equation is called mean-reverting jumped geometric Random Walk (MRJGRW) or mean-reverting jump diffusion with geometric Random Walk (MRJDwGRW). Additional possible Jumps are introduced as a sum of two random variables. First long-tailed *Q* is responsible for a promotional effect of album releasing (log-normal in our case). Second one *q* is obtained from unitary distribution and is related to first single, which is promoting whole album.

Releasing singles during record promotion is represented as a spike in sale (a random variable from the same distribution as first single: *q*) with mean duration between singles release: $1/p$. Singles appear as small spikes with a probability *p* during record promotion. Some records could be sold on a constant high level. We model it as an increase of the mean reverting value by varying a parameter *a(t)*. It is increased for short time by an additional random value *q* ( has the same distribution as for single's release). A general equation containing all effects:

$$dX_t = (a(t) - bX_t)dt + sdW_t + j(t)(dQ_t + dq_t) + i(t)dq_t \quad (5)$$

where:
if state 1−>2 , then *j(t)=1*, otherwise *j(t)=0*
if state=2 with probability *p* , then *i(t)=1*, otherwise *i(t)=0*
if state=2 and sub-state=2' , then *a(t)=a+q* , otherwise *a(t)=a*

    An unexpected popularity is implemented by additional sub-state 2'. Entering sub-state 2' (Popularity) means that record seems to touch a general audience (not only fans of that artist). This state can be reached from state 2 (Record promotion) only in points of single release, (*Probability*(2−>2')>0 only if *i(t)=1*). A transition matrix between the general state 2 and its sub-state 2', in this case there is no transition to base regime 1, can be calculated:

$$\{q_{ij}\}= \begin{bmatrix} q_{22} & q_{22'} \\ q_{2'2} & q_{2'2'} \end{bmatrix} = \begin{bmatrix} 1-q_{22'} & q_{22'} \\ q_{2'2} & 1-q_{2'2} \end{bmatrix} = \begin{bmatrix} 1-\text{pp}' & \text{pp}' \\ \text{p} & 1-\text{p} \end{bmatrix} \quad (6)$$

where *p* is a probability of a single release and *p'* is a probability of a success for a given single.
A system could be back in a normal state (1 or 2) with standard mean reverting *a(t)=a*, if next single is released (with the same probability *p*) or if record promotion is over (with probability *q21*). Thus, an average duration of staying in sub-state 2' (Popularity) is *q22/p*. Such an arbitrarily chosen methodology was introduced to imitate observed frequency of this state (unexpected popularity).

# 5 Model: Non-stationary case

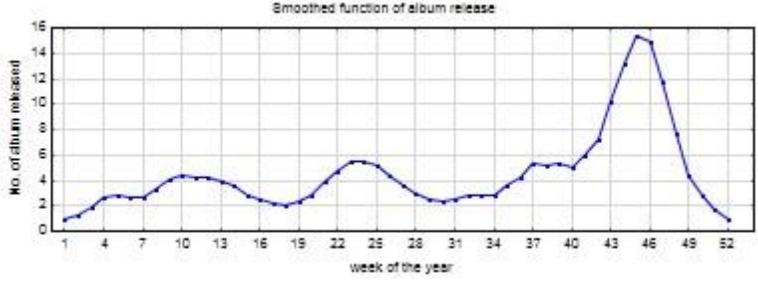

Figure 8: Album release seasonality as a function of probability correction

In addition to standard model, we introduce a seasonality time dependency of album release and the markovian size of a peak for a new release (size of such a peak is a function of a total sale of last previous album). Seasonality was introduced by rescaling probability of album release (element *q12* from transition matrix (4)). A probability of a new transition becomes now time dependent:

$$q_{12}(t_y) = q_{12} c(t_y) / E(c(t_y)) \qquad (7)$$

where *t* is a current week for a given year and correction function c(*t*) (Fig. 8) is normalized by an expected value of it *E(c(t))*.

The second innovation is a memory of sale. The height of the peak of sale on an album release date depends on total sale of the last previous album. We introduce a new random variable instead of log-normal Q from a stationary model:

$$Q_t(i) = \begin{cases} \text{for } i > 1 & \max(s_l(i-1)/s_c + s_s N(0,1), 0) \\ \text{for } i=1 & Q_t \end{cases} \qquad (8)$$

where *i* is an index of an album released by artist (first album peak height is the random variable like in stationary model), *sl(i)* is a total sale of previous album, *sc* is a scaling factor of that sale, *ss* is the level of noise.

# 6 Parameter estimation and fixing

Due to a big number of parameters only few numerical estimation procedures were applied and heuristic methods must supplement model's tuning. The unit of sale is [k=1000]. In the first wave values of matrix (4) were set up. Duration of promotion is defined as *1/q22=pE(N_singles)*=10*3.311=33.1 where *p* is a mean interval between singles (10 weeks) and *E(N_singles)* is an expected value of a number of singles per album. *E(N_singles)* was estimated based on histogram of observables (Fig. 7). An interval between an album release was chosen to be around the second peak of its empirical distribution *t*=131. Since an effective duration between two album releases in model follows *1/(q22+q12)*, so *q12*=97.9. In the second step mean-reverting geometric Random Walk (MRGRW), parameters were estimated for both the base and album promotion regime with known transition parameters by the maximal likelihood method and we get: *a*=1.16,*b*=267.512,*s*=0.25014. Unfortunately, due to the simplified model, short time series and threshold problem (long periods of sales below threshold with zeros instead), a significance of that estimation is roughly only. The peak values log-normal distribution parameters *Q=1000ln(N(u,s))*were estimated based on empirical observation and *u*=1.2, *s*=1.7. *q* was chosen arbitrary to be uniform in range (0,100). A success of entering a popularity regime appears in reality every 3 album on average so *1/3=p'E(N_singles), p'*=.1007. In nonstationary model some extra information was introduced. A correction function to seasonality of album release c(*t*) is just observed for albums distribution over weeks of the year smoothed by a moving average of frame 3 and the normalization factor *E(c(t))* is just an expected value of it. Such a correction is well known in statistics as a hazard function and have an important application in medicine. As the empirical seasonal dependence (hazard) is generally very erratic, it is customary to fit a smooth curve with moving average to enable the underlying shape to be seen [20]. In function of peak size based on the total size of previous album scaling factor *sc* and the level of noise *ss* were mean-square errors fits the best with an empirical distribution of sequential peaks of artist's sale.

# 7 Models comparison and system emergence

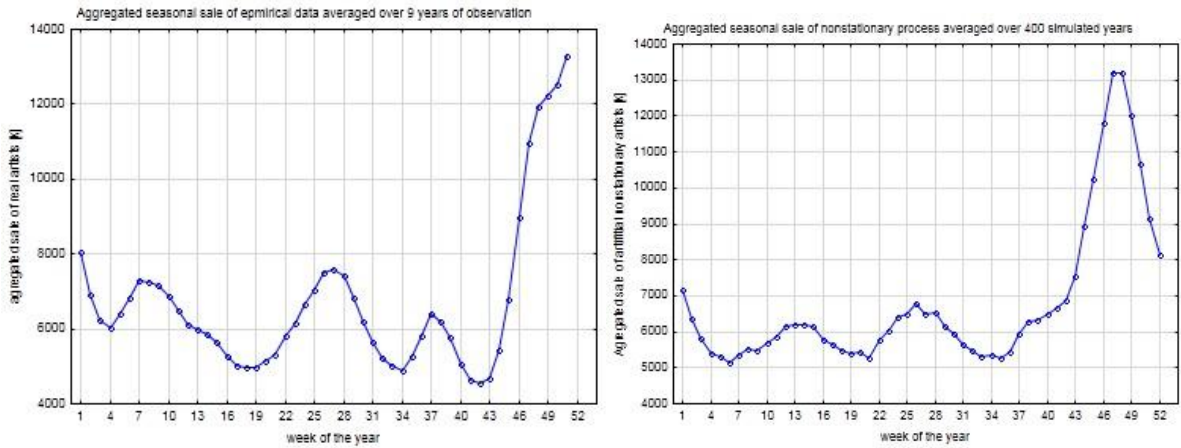

Figure 9: Smoothed (moving average with frame 3) trajectories of aggregated sales per week of the year. Empirical data [left], Non-stationary model [right]

The nonstationary model does not reproduce discontinuity observed between 52 and 1 week of the year in total sale (Fig. 9). Moreover, an implementation of seasonality of record sale (increase of album release late autumn and early winter) does not reflect the Christmas shopping madness. According to empirical data, the Christmas sales increase (Fig. 9) is different (difference between a standard level of sale and the Christmas peak is usually equal 300% for real market up to around 200% for a nonstationary model forming a triangle).

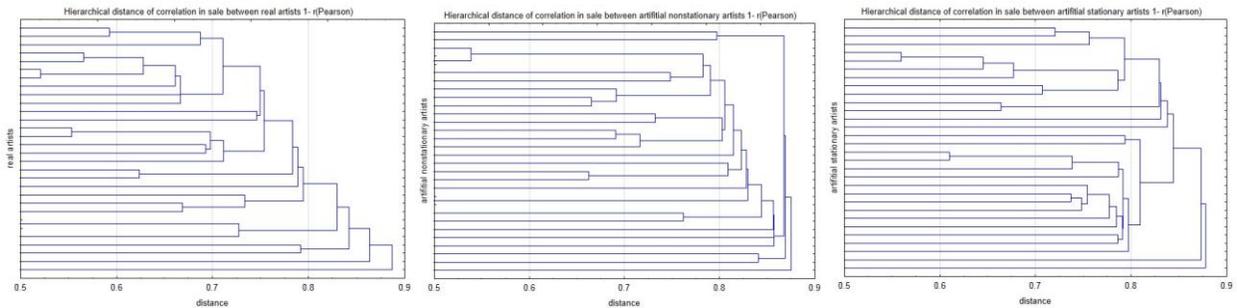

Figure 10: Hierarchical relations between artists: empirical data [left], Non-stationary model [Center], Stationary model [right]

Autumn's high intensity of celebrity premieres determines "Superstar sector" [1] on the hierarchical tree. We clam that high sales simultaneously recorded an influence on the correlation between the biggest stars, even if they belong to different music sectors. Both models do not reflect such clusters (Fig. 10) and do not differ significantly in terms of expected correlation coefficient. Nonstationarity is supposed to produce significantly higher correlation due to seasonality.

# 8  Conclusions

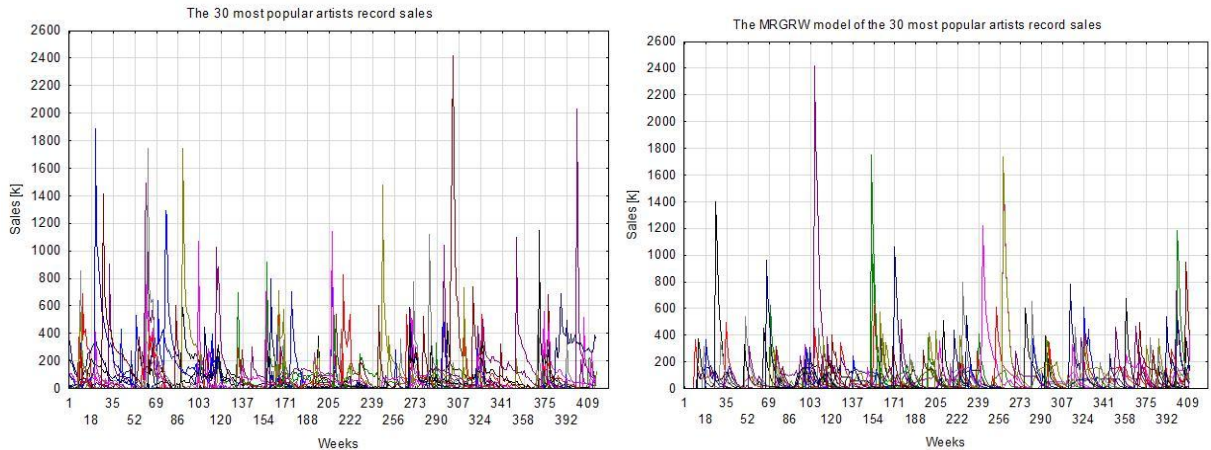

Figure 11: Visual comparison of Empirical data [left], Stationary model [right]

Provided models reflect the market behavior for best selling artist assuming their homogeneity (Fig. 11). A non-stationary mechanistic model replicates phenomenology of a collective behavior of the market (without Christmas time), although interaction terms were not introduced directly. Presented model framework is the first (up to our knowledge) publicly accessible attend to time series analyze of phonographic stardom sales (due to difficulty with data collection) and its great supplementary to many studies on chat performance (where data is widely available [5, 9, 10, 11]). A nonstationary model could be applied in sale prediction for the future record release for an individual artist and it could help in setting the premiere date. We also observe small disagreement between model and data, which are very interesting. Firstly, discontinuously in empirical total sale between 52 and 1 week of the year is not observed from the model simulation. Moreover, the relative peak of empirical sale during Christmas time is one order of magnitude higher than in the model. It means that customers wallets approach is crucial to understand that phenomena (after Christmas and New Year Eve people have not money for such a luxurious good like a record. We suggest, that the difference in a peak height could be understood as an activation on a new category of Christmas customers (introducing turbulence to the system), which suppose to be in a minority for the rest of the year. Such a newcomers are not following the 'rules' of phonographic market. Their shopping decisions seems not to relate to an actual album - product quality (significant difference between Christmas sale - Fig. 9) and they probably recognize only superstar artists - brands (a lack of a cluster of artists in the model - Fig. 10). On the other hand, non-Christian markets like Japanese one suit to our model because without the Christmas effect there is no big difference between Japanese record sales in January and December according to the real data.

# 9  Acknowledgments

We would like to express our sincere thanks to all our colleagues for fruitful discussions: Jaroslaw Kwapien, Joanna Janczura, Andrzej Grabowski, Tomasz Gubiec, Ryszard Kutner, Stanislaw Drozdz, Fredrik Liljeros and community of Summer Soliscate Conference. AJ thanks to Svenska Institutet for support.